\documentclass[aip,apl,reprint,twocolumn]{revtex4-1}

\usepackage{bm}%
\usepackage{graphicx}%
\usepackage{times}

\begin{document}


\title{Thermoelectric properties of ${\beta}$-FeSi$_{\text2}$}

\author{Tribhuwan Pandey$^{1}$, David J. Singh$^{2}$, David Parker$^{2}$ and Abhishek K. Singh$^{1}$}

\affiliation{$^{1}$Materials Research Centre, Indian Institute of Science, Bangalore
560012, India\\
  $^{2}$Materials Science and Technology Division, Oak Ridge National Laboratory, Oak Ridge, TN 37831-6056, USA }

\date{\today}

\begin{abstract}
We investigate the thermoelectric properties of ${\beta}$-FeSi$_{\text2}$ using first principles electronic structure and Boltzmann transport calculations. We report a high thermopower for both \textit{p}- and \textit{n}-type ${\beta}$-FeSi$_{\text2}$ over a wide range of carrier concentration and in addition find the performance for \textit{n}-type to be higher than for the \textit{p}-type. Our results indicate that, depending upon temperature, a doping level of 3$\times$10${^{20}}$ - 2$\times$10${^{21}}$ cm${^{-3}}$ may optimize the thermoelectric performance.

\end{abstract}

\maketitle

${\beta}$-FeSi$_{\text2}$ is a semiconductor~\cite{Dusausoy} of potential interest \cite{THEB} as a thermoelectric material~\cite{Ioffe,ware,Birkholz1,THEB}. In past studies it has been shown that ${\beta}$-FeSi$_{\text2}$ can be doped both as \textit{n}- and \textit{p}-type by substituting suitable elements on the iron and silicon sites~\cite{FeSi2_Co_1, FeSi2_Co_2, FeSi2_Mn, FeSi2_Mn2, FeSi2_P,FeSi2_Co}. Experimental study has reported ZT values of 0.2 and 0.4 for \textit{p}- and \textit{n}-type samples, respectively~\cite{ware,THEB}. Here ZT is defined as ZT = S{$^2$}${\sigma}$T/${\kappa}$, where S, ${\sigma}$ and ${\kappa}$ are thermopower, electrical, and thermal conductivity, respectively. Achieving high ZT values requires simultaneously high thermopower, high electrical conductivity and low thermal conductivity in the same material~\cite{Mahan}. In general, optimizing doping level is crucial for efficient thermoelectric performance~\cite{Mahan,Madsen_LiZnSb,Singh_PbTe,Shi,Parker_CrSi2} and it is not clear to what extent samples from existing studies have been optimized. In this present work, based on first principles electronic structure and Boltzmann transport theory calculations we address this issue.

These calculations were performed within density functional theory (DFT) using the general potential linearized augmented plane-wave (LAPW) method with local orbitals,\cite{singh_lo,singh_book} as implemented in the WIEN2k code~\cite{wien2k}. The LAPW sphere radii were 2.32 bohr and 2.06 bohr (1 bohr = 0.529177 \AA) for Fe and Si, respectively. In addition R\textit{k}$_{max}$ ${=}$ 9.0, is used  to ensure a well-converged basis set where, R and \textit{k}$_{max}$ are the smallest LAPW sphere radius and interstitial plane-wave cutoff, respectively. ${\beta}$-FeSi$_{\text2}$ crystallizes in an orthorhombic structure at temperatures below $\sim$ 1210K~\cite{tt}. The calculations for ${\beta}$-FeSi$_{\text2}$ were performed using the lattice parameters from Ref~\onlinecite{Dusausoy}. The internal atomic coordinates were determined by minimizing the total energy using the generalized gradient approximation (GGA) of Perdew and co-workers~\cite{PBE}. The electronic structure calculation were performed on a dense 18${\times}$18${\times}$18 ${\textbf{k}}$-point mesh. The transport calculations were done using Boltzmann transport theory~\cite{ziman} within the constant scattering time approximation (CSTA)~\cite{ziman,mermin}. The CSTA is based on the assumption that the scattering  time  does not vary strongly with energy. Note that the CSTA does not involve any assumptions regarding the usually strong temperature and doping level dependence of the scattering time. The advantage of this approximation is that the thermopower S(T) can be calculated without using any adjustable parameters. This has been used in the successful prediction of the thermoelectric properties of many materials~\cite{Parker_CrSi2,Singh_review}. We have used the BoltzTraP code~\cite{Madsen} for transport calculations on a dense 24${\times}$24${\times}$24 ${\textbf{k}}$-points mesh. 

\begin{figure}
\centering
\includegraphics[scale=0.45]{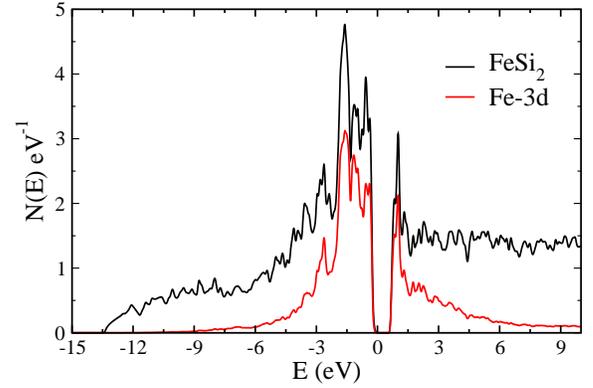}
\caption{Calculated total density of states and Fe 3\textit{d} projection for ${\beta}$-FeSi$_{\text2}$ on per formula unit basis.  We have chosen the energy zero as the valence band maximum.}
\label{fig:1}
\end{figure}

\begin{figure}
\centering
\includegraphics[scale=0.45]{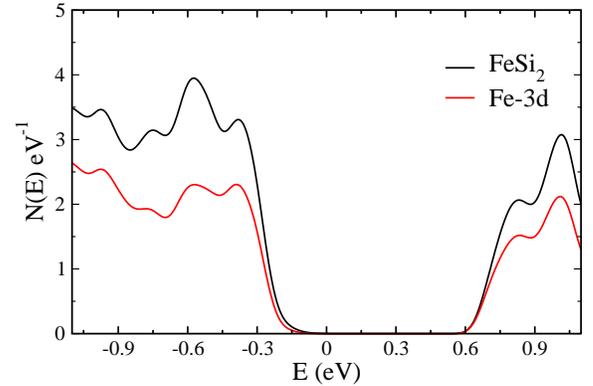}
\caption{Calculated total density of states near the band edges for ${\beta}$-FeSi$_{\text2}$. We have chosen the energy zero as the valence band maximum.}
\label{fig:2}
\end{figure}

The calculated electronic density of states (DOS) and Fe-3\textit{d} projection are shown in Fig.~\ref{fig:1}. Fig.~\ref{fig:2} depicts the DOS near the band edges.
The DOS shows the broad Si-\textit{p} bands merged with narrow Fe-\textit{d} bands producing a band gap of 0.67 eV. The valence band extends from about ${-}$13.5 eV below the valence band maximum (VBM). The lower part of the DOS from the bottom of the valence band to about ${-}$7 eV corresponds mainly to Si-\textit{s} states. Above this the valence and conduction both bands are derived from hybridized Si-\textit{p} and Fe-\textit{d} orbitals. There is an onset of heavy bands at $-$0.15 eV relative to the VBM, making the light band less important for the electrical transport. ${\beta}$-FeSi$_{\text2}$ has a very low DOS ranging from about $-$0.15 eV to 0, reflecting the energy range of the light band. This light band may negatively impact the \textit{p}-type performance in comparison to \textit{n}-type as in the case of skutterudite thermoelectrics~\cite{singh,yang}.

\begin{figure}
\centering
\includegraphics[scale=0.45]{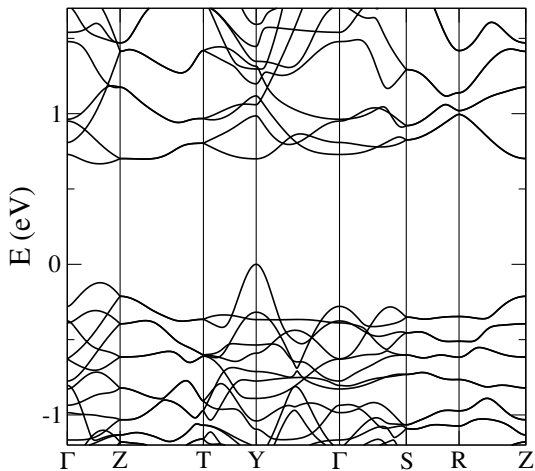}
\caption{The band structure of the  relaxed bulk structure of ${\beta}$-FeSi$_{\text2}$ is shown along high symmetry direction. We have chosen the energy zero as the valence band maximum.}
\label{fig:3}
\end{figure}

The band structure of ${\beta}$-FeSi$_{\text2}$ is shown in Fig.~\ref{fig:3}. The calculated band gap is indirect between Y (valence band) and a non-symmetry point between $\Gamma$ and Z (conduction band) with a value of 0.67 eV, which is in agreement with past theoretical studies~\cite{clark,antonov,Eisebitt}. For most materials DFT approaches tend to underestimate the band gap. However, for silicides the DFT band gap can be close to the experimental value.
As explained by Mattheiss~\cite{mattheiss1,matheiss}, this occurs since the hybridization effects which produce the band gap in silicides are well described by such calculations.

Experimental results on the band gap of ${\beta}$-FeSi$_{\text2}$ vary significantly. Photoluminescence emission, which indicates a direct gap transition, has been reported by a number of workers for ${\beta}$-FeSi$_{\text2}$~\cite{Arushanov,Hunt,regolini}. However, there are a few experimental reports suggesting an indirect transition. Filonov and co-workers~\cite{filonov} performed optical absorption measurements, which suggested an indirect band gap of 0.73 eV. Radermacher and co-workers~\cite{Radermacher} undertook photothermal deflection spectroscopy measurements,  finding an indirect band gap of 0.78 eV. Similarly Giannini et. al.~\cite{Giannini} reported an indirect band gap of 0.80 eV.  Further band gap experimental investigation would be helpful.

Returning to the details of the band structure in Fig.~\ref{fig:3} , the conduction bands are heavy and anisotropic. They are also substantially non parabolic even at relatively low energies above the conduction band minimum (CBM). The VBM is derived from a single light band of Si character as reported earlier. As shown in the plot the conduction band contains three critical points at Y, ${\Gamma}$, and Z at 0.03, 0.06 and 0.035 eV above the CBM, respectively. These pockets will be accessible at temperatures of room temperature and above giving rise to effective degeneracy and multi-carrier transport.
\begin{figure}[h!]
\centering
\includegraphics[scale=0.45]{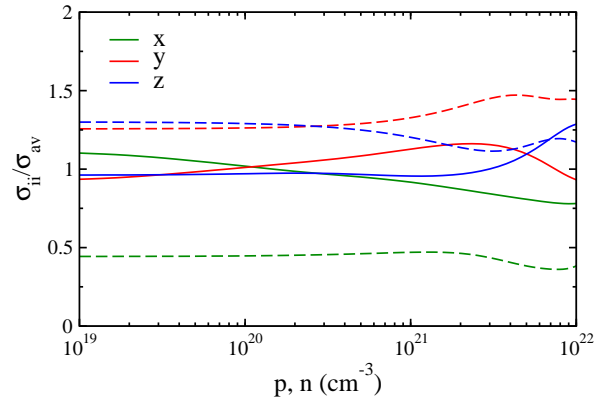}
\caption{Conductivity anisotropy as a function of carrier concentration for p-(solid lines) and n-type (dashed lines) FeSi$_{\text2}$ at 300K. }
\label{fig:4}
\end{figure}

CSTA yields the conductivity only up to a multiplicative constant of an average relaxation time ${\tau}$. However the conductivity anisotropy $\sigma_{ii}$/$\sigma_{av}$ can be calculated without knowing ${\tau}$ (where $\sigma_{av}$ is $\sigma_{av}$ = ($\sigma_{xx}$+$\sigma_{yy}$+$\sigma_{zz}$)/3). As discussed above the bands comprising the CBM and VBM have significant anisotropy, suggesting an anisotropy in the conductivity.  As shown in Fig.~\ref{fig:4}, the high conductivity direction is different for \textit{p}-type doping (x) and \textit{n}-type doping (z) for this material at 300K. This is unusual because normally the conductivity reflects the bonding properties of a material, which are independent of carrier type. This same unusual effect can be seen is some other silicides as well e.g CrSi$_{\text2}$~\cite{Parker_CrSi2}.

\begin{figure}[h!]
\centering
\includegraphics[scale=0.45]{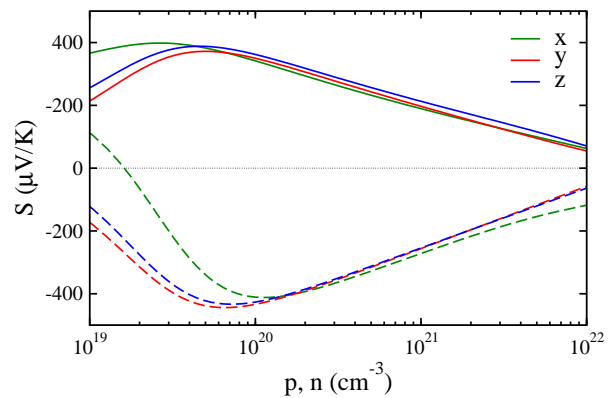}
\caption{Thermopower along different directions as a function of carrier concentration for p-(solid lines) and n-type (dashed lines) FeSi$_{\text2}$ at 900K.}
\label{fig:5}
\end{figure}

For an anisotropic semiconductor, the Mott formula suggests that thermopower remains isotropic in the limit of high carrier concentration and low temperature regardless of the details of the effective mass or its anisotropy.  Thus the thermopower is normally more isotropic than the conductivity  in semiconductors.  We present the directionally dependent 900 K thermopower in Fig~\ref{fig:5}. 

The results show a large anisotropy in the thermopower at low concentration for both \textit{p}- and \textit{n}-type carriers. However, at high carrier concentration the thermopower shows little anisotropy. The low carrier concentration thermopower anisotropy comes from large and different anisotropy of the conductivity for \textit{p}- and \textit{n}-type, as described above. Interestingly, the low conductivity directions for \textit{n}-type (x and z) are the high conductivity direction for \textit{p}-type giving rise to a strong bipolar effect along these directions. This is most prominent for the x direction which is
the low conductivity direction for \textit{n}-type and is the highest conductivity direction for \textit{p}-type. This bipolar effect can be clearly seen in Fig.~\ref{fig:5}. The anisotropy at high carrier concentration comes from the non-parabolic bands.

\begin{figure}[h!]
\centering
\includegraphics[scale=0.45]{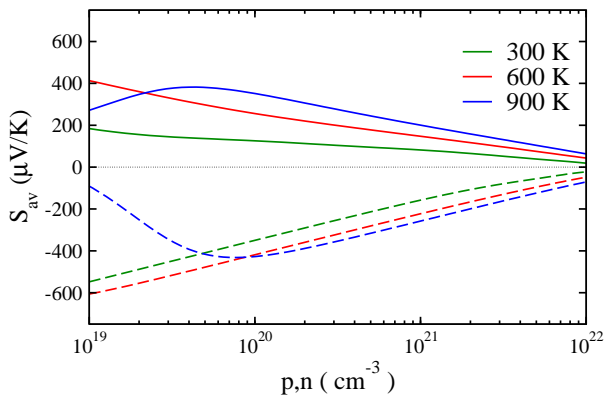}
\caption{Direction averaged thermopower as a function of carrier concentration for \textit{p}- (solid lines) and \textit{n}-type (dashed-lines) FeSi$_{\text2}$.}
\label{fig:6}
\end{figure}

Fig.~\ref{fig:6} shows the direction averaged thermopower as a function of carrier concentration. The average thermopower is calculated by using the S$_{av}$ = ($\sigma_{xx}$S$_{xx}$+$\sigma_{yy}$S$_{yy}$+$\sigma_{zz}$S$_{zz}$)/(3$\sigma_{av}$), where $\sigma_{av}$ can be given as $\sigma_{av}$ = ($\sigma_{xx}$+$\sigma_{yy}$+$\sigma_{zz}$)/3. On comparing these results with the directionally dependent thermopower Fig.~\ref{fig:5} one may note that the bipolar effect is still quite prominent. This is because the high conductivity direction dominates the conductivity average, which is the direction with weakest bipolar reduction.

To understand the qualitative effect of transport coefficients on the thermoelectric properties, we recall the expression for figure of merit ZT = S{$^2$}${\sigma}$T/${\kappa}$, where thermal conductivity ${\kappa}$ is ${\kappa}$ = ${\kappa_{e}}$ + ${\kappa_{l}}$, i.e. the sum of electronic thermal conductivity (${\kappa_{e}}$) and lattice thermal conductivity (${\kappa_{l}}$). It is convenient to rewrite this as ZT = \textit{r}S{$^2$}/L, \textit{r} being the ${\kappa_{e}}$/${\kappa}$, with the obvious restriction \textit{r} $\leq$ 1 and L = ${\kappa_{e}}$/${\sigma}$T, the Lorenz number as defined by the Wiedemann-Franz relation. This relation shows S as the main factor in determining ZT. This equation explains that bipolar conduction has two main negative effects on thermoelectric performance:  firstly,  it reduces thermopower (S) and secondly it increases L. Finally, this formula shows the importance of the lattice thermal conductivity in setting the limit for the achievable ZT. The thermopower of a semiconductor generally can be enhanced by lowering the carrier concentration up until the onset of bipolar conduction (where the thermopower begins to decrease with decreasing carrier concentration, the opposite of the usual situation). However, for most materials optimum performance typically occurs at carrier concentrations above those where the maximum S occurs.

We present in Fig.~\ref{fig:6} the calculated thermopower for \textit{p}-type at 900K.  It shows unipolar behavior over a wide range of carrier concentration.   Only for carrier concentrations less than $\sim$ 6$\times$10${^{19}}$ cm${^{-3}}$ does the bipolar effect come into play. At higher carrier concentrations such as 10${^{21}}$ cm$^{-3}$ thermopower still has magnitudes greater than 200$\mu$V$/$K. At temperatures lower than 900K, i.e. at 300 and 600K the thermopower is largely free from bipolar effects in the investigated carrier concentration range. For these temperatures  higher thermopower can be achieved at lower carrier concentration (typically, 1$\times$10${^{19}}$ cm${^{-3}}$ - 2$\times$10${^{19}}$ cm${^{-3}}$), but at such low carrier concentrations and temperatures the dominance of the lattice thermal conductivity will suppress the performance. 

For \textit{n}-type samples the situation is rather different. At 900K bipolar effects are significant when the carrier concentration is less than 9$\times$10${^{19}}$ cm${^{-3}}$. As discussed above due to the absence of light band and presence of multi-carrier transport \textit{n}-type leads to higher thermopower than \textit{p}-type even at high carrier concentrations. This is in accordance with previous experimental studies~\cite{ware,THEB}, where for $n$-type samples higher ZT has been reported. At temperatures lower than 900K the thermopower is generally not impacted by bipolar conduction in the investigated carrier concentration range. Furthermore for \textit{n}-type thermopower has magnitudes higher than 200$\mu$V$/$K over a wide range of carrier concentration. For efficient thermoelectric performance a high thermopower magnitude (typically 200$\mu$V$/$K - 300$\mu$V$/$K) free from bipolar effects is desirable. We report for $n$-type ${\beta}$-FeSi$_{\text2}$ that this can be achieved in the carrier concentration range of 3$\times$10${^{20}}$ - 2$\times$10${^{21}}$ cm${^{-3}}$ depending upon temperature.

In conclusion, Boltzmann transport theory calculations for ${\beta}$-FeSi$_{\text2}$ show a large anisotropy in electrical conductivity and thermopower at lower carrier concentrations for both \textit{p}- and \textit{n}-type carriers. The thermopower results suggest that ${\beta}$-FeSi$_{\text2}$ should be investigated for thermoelectric performance for the electron doping of 3$\times$10${^{20}}$ - 2$\times$10${^{21}}$ cm${^{-3}}$, which roughly corresponds to 0.09-0.61 electron/unit-cell.  While we do not calculate ZT, we predict high thermopower at high carrier concentration, which is observed in many high performance thermoelectrics. Achieving such a high doping concentration may be a challenge. In transition metal disilicides doping on the metal site (Fe in this case) is less favorable because of a higher contribution from Fe \textit{d} states near the band edges as shown in DOS plot.  For this reason a more likely site for doping would be the Si and interstitial sites.  In particular, As and P doping on the Si site may be indicated.

{\bf Acknowledgments} T. P. acknowledges the hospitality of the Oak Ridge National Laboratory.  Work funded by the U.S. Department of Energy, Office of Science, Basic Energy Sciences, Materials Sciences and Engineering Division (T.P., D.J.S, first principles). and the D.O.E. Energy Efficiency and Renewable Energy, Vehicle Technologies program (D.P., analysis).

\end{document}